\documentclass[10pt]{article}
 \usepackage[dvips]{graphicx}
 \usepackage{amsmath}
 \usepackage{amsxtra}
\begin{document}
\title {Magueijo-Smolin Transformation as a Consequence of a Specific Definition of Mass,Velocity, and  the Upper limit on Energy}
\bigskip
\author{ Alex Granik\thanks{Department of
Physics, University of the
Pacific,Stockton,CA.95211;~E-mail:~agranik@pacific.edu}}
\date{}
\maketitle
\begin{abstract}
We consider an alternative approach to  double
special-relativistic theories. The point of departure is not
$\kappa$-deformed algebra (or even group-theoretical
considerations) but rather 3 physical postulates defining
particle's velocity, mass, and the upper bound on its energy in
terms of the respective classical quantities. For a specific
definition of particle's velocity we obtain Magueijo-Smolin (MS)
version of the double special-relativistic theory. It is shown
that this version follows from the $\kappa$-Poincare algebra by
the appropriate choice of  on the shell mass , such that it is
always less or equal Planck's mass. The $\kappa$-deformed
Hamiltonian is found which invalidates the recent arguments about
unphysical predictions of the MS transformation.
\end{abstract}

A recent research
(e.g.\cite{JK1},\cite{JL1},\cite{JL2},\cite{JR1},\cite{GA1},\cite{
GA2},\cite{MS}) on the so-called double special relativity  not
only reexamined its relation to $\kappa$-deformed kinematics, but
in one specific example \cite{JR1} also subjected to criticism
physical predictions of one of these theoretical constructs \cite{MS}.\\

It should be mentioned that as early as in $1994$, J.Lukierski
with collaborators \cite{PK} demonstrated that there exist an
infinite set of transformations reducing the $\kappa$-deformed
Casimir in Majid-Ruegg basis \cite{MR} (used in all the
double-special relativistic theories, e.g.\cite{NB}) to the
diagonal form. In fact, it is possible to show that any of these
transformations correspond to a different choice of what one can
consider as a definition of the deformed mass. This makes it
difficult, without any additional assumptions, to choose a unique
physical theory corresponding to the respective transformation.
This difficulty is emphasized \cite{JR1} by what looks like
apparent non-physical predictions of one of these constructions
\cite{MS}.\\

Here we revisit the latter work \cite{MS}, more specifically its
treatment of the energy-momentum domain, this time departing not
from the group-theoretical point of view, but rather from certain
physically justified restrictions (postulates) imposed on the
classically-defined physical quantities, namely energy, mass, and
velocity. An analogous approach was used in \cite{JL2} for a more
narrowly defined goal: a study of possible definitions of
$\kappa$-deformed velocities and their addition laws.

We begin by introducing the postulates defining

i) the velocity of a particle,

ii)its mass to be the same for any scale( from classical to Planck
scale) and therefore based on the relations provided by the
momentum sector of classical relativity,

iii)the existence of the upper bound (the Planck energy) on the
values of both energy and momentum.\

We also retain the upper bound ( speed of light $c$) on a particle
velocity.\\

In what follows we use units where $c=1$, Planck constant
$\hbar=1$, and Boltzmann constant $k$=1. We denote the Planck
energy (momentum) by $\kappa$ which in these units is equal to the
inverse of the Planck length $\lambda$ ($\kappa=1/\lambda$). The
classical relation between energy $P_0$ and momentum $P_i$ in
these units has the following dimensionless form:
\begin{equation}
\label{1} \Pi_0^2-|\Pi|^2=\mu^2
\end{equation}
where

\begin{center}$\mu\equiv m/\kappa,~~ \Pi_i=P_i/\kappa, ~~\Pi_0=P_0/\kappa, ~~
i=1,2,3$.\end{center} and $m$ is particle's mass.\\

Quite analogously we introduce the dimensionless expressions for
the physical energy $p_0$ and momentum $p$ (different from the
above energy $\Pi_0$ and momentum $\Pi_i$) applicable in the
region of Planck-scale physics

\begin{center}$\pi_0=p_0/\kappa,~~ \pi_i=p_i/\kappa,~~ i=1,2,3$
\end{center}

Following \cite{JL1} we write the general functional relation
between the classical energy-momentum $\Pi_{\alpha}, \alpha
=0,1,2,3$ ( not physical anymore in the Planck-scale phenomena)
and its Planck-scale counterpart $\pi_{\alpha},\alpha =0,1,2,3$:
\begin{equation}
\label{2} \pi_0=f(\Pi_0),  \pi_i=g(\Pi_0)\Pi_i
\end{equation}
where the functions $f(\Pi_0)$ and $g(\Pi_0)$ to be defined.
\\

To find these functions we use the above postulates (i)-(iii). The
dimensionless velocity of a particle $v \leq 1$ (compatible with
its classical definition in terms of the energy-momentum) is
defined as follows
\begin{subequations}
\begin{eqnarray}
\label{3a} v^2=(\frac{\Pi}{\Pi_0})^2=(\frac{\pi}{\pi_0})^2;\nonumber\\
\pi^2=\sum_{i=1}^3\pi_i\pi_i,\hspace{0.1in} \Pi^2=\sum_{i=1}^3\Pi_i\Pi_i 
\end{eqnarray}

where $p^2=p_ip^i$. Note that in this definition the velocity
$v_i$ looks as the one used in the classical case, except that now
this velocity
$$v^2\neq(\partial{\pi_0}/\partial{\pi})^2,$$ while in the classical
case

\begin{equation}
\label{3b}
v^2=(\frac{\Pi}{\Pi_0})^2=(\frac{\partial{\Pi_0}}{\partial{\Pi}})^2
\end{equation}
\end{subequations}

Next we use the second postulate (ii) which defines particle's
mass $\mu$ to be {\it the same} in {\it all the regions} (from
classical to Planck scale), and independent of the velocity
definition (\ref{3a})-(\ref{3b}).
\begin{equation}
\label{4}\frac{1}{\mu}=2\lim_{\Pi^2\rightarrow 0}\frac{\partial
\Pi_0}{\partial \Pi^2}=2\lim_{\pi^2\rightarrow 0}\frac{\partial
\pi_0}{\partial \pi^2}
\end{equation}

Finally we require ( postulate iii) that
\begin{equation}
\label{5} |\pi|\leq 1,~~ |\pi_0|\leq 1
\end{equation}
where the equality signs correspond to $\Pi_0,|\Pi|\rightarrow
\infty$\\

We begin with the velocity definition according to Eq. (\ref{3a}).
Upon substitution of this equation into Eq.(\ref{2}) we obtain
\begin{equation}
\label{6} f(\Pi_0)=g(\Pi_0)\Pi_0
\end{equation}
This means that
\begin{eqnarray}
\label{7} 
\pi_0=g(\Pi_0)\Pi_0,\nonumber\\
d\pi_0=\Pi_0\frac{dg}{d\Pi_0}d\Pi_0+g(\Pi_0)d\Pi_0,\nonumber\\
d\pi^2=\Pi^2\frac{d g^2}{d\Pi_0}d\Pi_0+g^2d\Pi^2
\end{eqnarray} 
Inserting Eq.(\ref{7}) into the definition of mass Eq.(\ref{4}) we
arrive at the following differential equation:
\begin{equation}
\label{8} \Pi_0\frac{dg}{d\Pi_0}+g(\Pi_0)-[g(\Pi_0)]^2=0
\end{equation}
Its solution is:
\begin{equation}
\label{9} g(\Pi_0)= \frac{1}{1\mp A\Pi_0}
\end{equation}
where the integration constant$A$ to be determined on the basis
of the above postulates.\\

As a result, according to Eqs.(\ref{2}),(\ref{6}) the
energy-momentum $\pi_\alpha (\alpha=0,1,2,3)$ is:
\begin{subequations}
\begin{equation}
\label{10a} \pi_i=\frac{\Pi_i}{1\mp A\Pi_0}
\end{equation}
\begin{equation}
\label{10b} \pi_0=\frac{\Pi_0}{1\mp A\Pi_0}
\end{equation}
\end{subequations}

The value of the integration constant $A$ and the choice of the
respective sign in the obtained solution (\ref{10a}),(\ref{10b})
are dictated by our postulate (iii), Eq. (\ref{5}).\\

To determine both, we notice that since in the classical limit
$\pi_{\alpha} \rightarrow \Pi_{\alpha}$ the positive(negative)
 values of $\Pi_0$ should correspond to positive (negative) values
of $\pi_0$ respectively. This means the following:
\begin{subequations}
\begin{equation}
\label{11a}
 \pi_0=
\frac{\Pi_0}{1+ A\Pi_0}, ~~~~{\Pi_0> 0,\pi_0 > 0}
\end{equation}
\begin{equation}
\label{11b}
 \pi_0=\frac{\Pi_0}{1-A\Pi_0},~~~~~{\Pi_0< 0,\pi_0 < 0}
\end{equation}
\end{subequations}

Taking the limit $\Pi_0 \rightarrow +\infty(-\infty)$ of
Eq.(\ref{11a}),(Eq. \ref{11b}) and using our postulate iii)
(Eq.\ref{5}) we get
\begin{equation}
\label{12} A=1
\end{equation}

Inserting this value of $A$ into
Eqs.(\ref{10a}),(\ref{11a}),(\ref{11b}) we obtain the explicit
expressions for $\pi_i$ and $\pi_0$
\begin{eqnarray}
\label{13} \pi_0=\frac{\Pi_0}{1+\Pi_0}&&\Pi_0,\pi_0>
0\nonumber\\
\pi_0=\frac{\Pi_0}{1-\Pi_0}&&\Pi_0,\pi_0< 0
\end{eqnarray}

\begin{eqnarray}
\label{14} \pi_i=\frac{\Pi_i}{1+\Pi_0}&&\Pi_0,\pi_0>
0\nonumber\\
\pi_i=\frac{\Pi_i}{1-\Pi_0}&&\Pi_0,\pi_0< 0
\end{eqnarray}

These expressions reproduce the results obtained in \cite{MS} with
the only difference that here $\pi_0$ is the antisymmetric
function of $\Pi_0$ in contradistinction to \cite{MS}.\\

If we use the classical expressions for $\Pi_{\alpha},~ \alpha
=0,1,2,3$ ( with the same $v_i$ and $\mu$ in all the regions)
$$\Pi_0=\mu\gamma, ~~\Pi_i=\mu v_i\gamma,~~~~
\gamma=\frac{1}{\sqrt{1-v^2}}$$ and Eqs.(\ref{10a}),(\ref{10b})
then we readily obtain ( restricting our attention to the positive
region of $\pi_0$) the respective expressions (cf.\cite{MS}) for
$\pi_{\alpha}$:
\begin{eqnarray}
\label{add} \pi_0=\frac{\mu\gamma}{1+\mu\gamma}\nonumber \\
\pi_i=\frac{\mu v_i\gamma}{1+\mu\gamma}
\end{eqnarray}
From (\ref{add}) follows that  the rest energy $\pi_0^0$ less than
the mass $\mu$:
\begin{equation}\label{add2}
\pi_0^0=\frac{\mu}{1+\mu}\leq \mu
\end{equation}
Here the equality sign corresponds to the classical region $\mu
\equiv m/\kappa << 1$\\

From expressions (\ref{13}), (\ref{14}) we obtain the inversion
formulas:
\begin{eqnarray}
\label{15} \Pi_0=\frac{\pi_0}{1-\pi_0}&&\Pi_0,\pi_0>
0\nonumber\\
\Pi_0=\frac{\pi_0}{1+\pi_0}&&\Pi_0,\pi_0< 0
\end{eqnarray}
\begin{eqnarray}
\label{16} \Pi_i=\frac{\pi_i}{1-\pi_0}&&\Pi_0,\pi_0>
0\nonumber\\
\Pi_i=\frac{\pi_i}{1+\pi_0}&&\Pi_0,\pi_0< 0
\end{eqnarray}

If we use  classical Casimir and expressions (\ref{15}) and
(\ref{16}) then the respective Casimirs for the energy-momentum in
the Planck region are
\begin{subequations}
\begin{equation}
\label{17a} \frac{\pi_0^2}{(1- \pi_0)^2}-\frac{\pi^2}{(1-
\pi_0)^2}=\mu^2
\end{equation}
\begin{equation}\label{17b}
\frac{\pi_0^2}{(1+ \pi_0)^2}-\frac{\pi^2}{(1+ \pi_0)^2}=\mu^2
\end{equation}
\end{subequations}

where ($-$) corresponds to the positive values of $\pi_0$ and
($+$) corresponds to negative values of $\pi_0$. Solving
Eqs.(\ref{17a}),(\ref{17b}) with respect to $\pi_0$ and choosing
the correct signs ( according to the positive and negative values
of $\pi_0$, remembering that both $|\pi|,|\pi_0|\le 1$) we arrive
at the following relation
\begin{figure}
 \begin{center}
 \includegraphics[width=6cm, height=6cm]{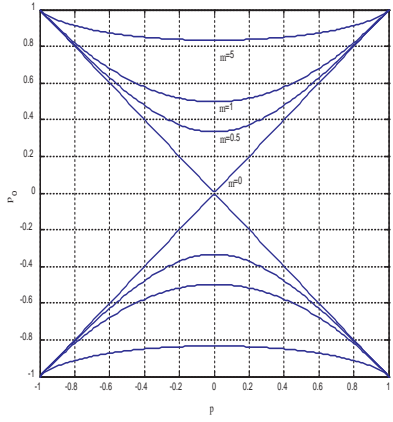}
 \caption{\small  dispersion relation $\pi_0=F(\pi)$;
 $4$ curves correspond to the values of masses $\mu=0,0.5,1.0,
 5.0$}
 \end{center}
 \end{figure}

\begin{equation}
\label{18}
\pi_0=\pm\frac{\sqrt{\pi^2(1-\mu^2)+\mu^2}-\mu^2}{1-\mu^2}
\end{equation}
where the upper(lower) sign corresponds to $\pi_0>0 (\pi_0<0)$
respectively.  It is seen that the regions of the positive and
negative values of $\pi_0 =F(\pi)$ are the same with accuracy to
the sign. The graph of $\pi_0 =F(\pi)$ is shown in Fig.1.\\

Based on the relation between $\pi_{\alpha}$ and $\Pi_{\alpha}$
[Eqs. (\ref{11a}), (\ref{11b}), (\ref{13}),(\ref{14})] and on  the
expressions for classical Lorentz boost of $\Pi_{\alpha}$ in the
$z$-direction with a velocity $V_3$ (in the units of $c =1$), we
can calculate in an elementary fashion the respective boost
relations ( found by Magueijo and Smolin in \cite{MS} with the
help of group-theoretical analysis ) energy- momentum
$\pi_{\alpha}$ at the Planck scale . We write them in the
dimensionless form:
\begin{subequations}
\begin{equation}
\label{19a}
\pi_0'=\frac{\Gamma(\pi_0-V_3\pi_3)}{1-\pi_0+\Gamma(\pi_0-V_3\pi_3)}
\end{equation}
\begin{equation}
\label{19b}
\pi_i'=\frac{\delta_{3i}\Gamma(\pi_i-V_3\pi_0)+\pi_i(1-
\delta_{3i})}{1-\pi_0+\Gamma(\pi_0-V_3\pi_3)}, ~~~~~~ i=1,2,3
\end{equation}
\end{subequations}
Here $\delta_{3i}$ is the Kroenecker delta-function and
$\Gamma=1/\sqrt{1-V_3^2}$. Since  particle's velocity has been
defined as $v_i=\pi_i/\pi_0$, it is not surprising that
Eqs.(\ref{19a},\ref{19b}) yield the velocity addition rule ,
coinciding with the classical relativistic rule:
\begin{equation}
v_i'=\frac{\delta_{3i}(v_i-V_3)}{1-v_3V_3}+\frac{v_i(1-\delta_{3i})}
{\Gamma(1-v_3V_3)}
\end{equation}

It was shown in  \cite{JK1} that within the context of
$\kappa$-Poinciana algebra various possible doubly-special
relativity constructions can be viewed as different bases of this
algebra. In particular, Magueijo-Smolin basis \cite{MS} is one of
such bases. In a more general scheme of things , J.Lukierski and
collaborators \cite{PK} demonstrated that all possible double
special relativistic constructions differ by a
suitable choice of what one defines as an effective mass.\\

Here we show that the Magueijo-Smolin transformation (obtained
here in an elementary fashion with the help of simple physical
postulates)contains an additional physical constraint on the mass.
This transformation can be derived pro-forma from
$\kappa$-deformed algebra, and the result explicitly shows that in
this case the particle mass $\mu \equiv m/\kappa \le 1$. To
demonstrate that we write the relations of the classical basis
(denoted here as $\overline{\Pi}_{\alpha},~\alpha=0,1,2,3$) by
rearranging the formulas given in \cite{PK}:

\begin{equation}
\label{20a}
\overline{\Pi}_0=A[e^{\overline{\pi}_0}-cosh(\overline{\mu})]\equiv
Ae^{\overline{\pi}_0}[1-e^{-\overline{\pi}_0}cosh(\overline{\mu})]
\end{equation}
\begin{equation}
\label{20b}
\overline{\Pi}_i=Ae^{\overline{\pi}_0}\overline{\pi}_i
\end{equation}

where $A$ is an arbitrary constant to be determined. The
respective Casimirs  are
\begin{equation}
\label{21}
cosh(\overline{\pi}_0)- \frac{\pi^2
e^{\overline{\pi}_0}}{2}=cosh(\overline{\mu})
\end{equation}
and
\begin{equation}\label{22}
\overline{\Pi}_0^2-\overline{\Pi}^2= A^2sinh^2(\overline{\mu})
\end{equation}\\

We can symmetrize expressions (\ref{20a}), (\ref{20b}) by
introducing the following quantities:

\begin{subequations}
\begin{equation}
\label{23a}
  \pi_0=e^{-\overline{\pi}_0}\frac{\overline{\Pi}_0}{A}=
  [1-e^{-\overline{\pi}_0}cosh(\overline{\mu})]
\end{equation}
\begin{equation}
\label{23b}
\pi_i=\overline{\pi}_i
\end{equation}
\end{subequations}
Combining  (\ref{23a}),  (\ref{23b}) and (\ref{20a}), (\ref{20b}),
we obtain:
\begin{subequations}
\begin{equation}
\label{24a} \overline{\Pi}_0=
Acosh(\overline{\mu})\frac{\pi_0}{1-\pi_0}
\end{equation}
\begin{equation}\label{24b}
\overline{\Pi}_i=Acosh(\overline{\mu})\frac{\pi_i}{1-\pi_0}
\end{equation}
\end{subequations}
By comparing Eqs.(\ref{24a}) and (\ref{24b}) with Eqs. (\ref{13})
and (\ref{14}) we immediately see that Magueijo-Smolin basis
follows from $\kappa$-deformed algebra if and only if
\begin{equation}
\label{25}
A= \frac{1}{cosh(\overline{\mu})}
\end{equation}
In this case Casimir (\ref{22}) reads
\begin{equation}\label{26}
  (\frac{\pi_0}{1-\pi_0})^2-(\frac{\pi}{1-\pi_0})^2=(tanh^2(\overline{\mu})
\end{equation}
Comparing Eq.(\ref{26}) with Casimir given by Eq.(\ref{17a})
\cite{MS} we arrive at the conclusion that the mass $\mu$ used in
the latter is
\begin{equation}\label{27}
\mu=tanh(\overline{\mu}) \le 1, ~~~~~~~~~~~Q.E.D
\end{equation}

In addition, if we use (\ref{add2}) then Eq.(\ref{27}) imposes the
following condition on the value of the rest energy
$\pi_0|_{|pi|=0} \equiv \pi_0^0$:
\begin{equation}\label{28}
\pi_0^0 =\frac{1}{2}
\end{equation}\\

At the first glance this condition looks ( predicated on the
restriction on particle's mass $\mu \le 1$) as overly restrictive.
On the other hand, considering Magueijo-Smolin transform as a
"free-standing" transformations we are not forced to have an upper
bound on the rest energy $\pi_0^0$  in the Planck region where
$\pi_0^0 =1/2$ ( that is one half of the upper bound on $\pi_0$).
Still, there is a strong argument in favor of adopting the upper
bound on particle's mass. If we would like to be consistent then
it seems quite reasonable to expect that all the quantities in the
region of Planck scales to be bounded from above by Planck energy
$\kappa$, or bounded from below by Planck length $\lambda$.\\

We have derived Magueijo-Smolin transformation without resorting
to group-theoretical approach by simply defining particle velocity
$v_i$, its mass $\mu$ , and the upper bound on the magnitude of
momentum-energy $\pi_{\alpha}$ in terms of the respective
classical quantities. In particular, the velocity $v_i$ is defined
as $v_i=\pi_i/\pi_0$ ( Eq.\ref{3a}). If we substitute into this
definition the relations between $\pi_i,\pi_0$ and the respective
quantities $\overline{\pi}_i, \overline{\pi}_0$, Eqs. (\ref{23a}),
(\ref{23b}) (used in the conventional treatment based on
$\kappa$-Poincare algebra) we arrive at the value of the velocity
$v_i$ which is exactly the right group velocity $V^R_i$ (obeying
classical addition law) introduced in \cite{JL2}:
\begin{equation}\label{29}
 v_i= V^R_i=
 \frac{{e^{\overline{\pi}_0}\overline{\pi}_i}}{e^{\overline{\pi}_0}-cosh(\overline{\mu})}
\end{equation}
Interestingly enough, the particle velocity is identical in 2
different bases: Magueijo-Smolin basis and the basis used in
\cite{JL2}.\\

To complete our elementary treatment of Magueijo-Smolin transform,
we address its critique expressed in \cite{JR1}. It is argued
there that a definition of the particle velocity according to
Hamilton equations results in a paradoxical situation where $2$
particles of different masses moving in an inertial frame with the
same velocity will have different velocities when viewed from
another inertial frame. The fallacy of this conclusion is due to
the fact that the authors of \cite{JR1} used a $non$-$deformed$
hamiltonian
formalism.\\

It has been demonstrated (e.g.,\cite{JL3},\cite{JK2}) that a
velocity definition in a non-commutative space is dictated by an
appropriate choice of a $deformed$ Hamiltonian formalism ( see
also \cite{JL2}). To this end let us consider dimensionless
relativistic phase space variables
$Y_A=(\xi_{\alpha},\pi_{\alpha}),
(\xi_{\alpha}=x_{\alpha}/\lambda, \alpha=0,1,2,3)$ normalized by
the appropriate Planck scales and whose commutation relations are:

\begin{eqnarray}
\label{30}
[\pi_i,\xi_j]&=&\delta_{ij}, \nonumber \\
\lbrack\pi_0,\xi_0\rbrack &=& -(1-\pi_0), \nonumber \\
\lbrack\pi_0,\xi_i\rbrack &=&0, \nonumber \\
\lbrack \xi_0,\xi_i\rbrack &=&\xi_i~ ,  \nonumber \\
\lbrack \xi_0,\pi_i\rbrack &=& -\pi_i, \nonumber \\
\lbrack\pi_{\alpha},\pi_{\nu}\rbrack &=&0
\end{eqnarray}
The $\kappa$-deformed Hamilton equations then yield
(cf.\cite{JL2}):
\begin{subequations}
\begin{equation}\label{31a}
  \frac{d\xi_i}{ds}=-\frac{\partial H^{(\kappa)}}{\partial \pi_i},
\end{equation}
\begin{equation}\label{31b}
  \frac{d\xi_0}{ds}=-\sum_1^3\pi_i\frac{\partial H^{(\kappa)}}{\partial \pi_i}+
  (1-\pi_0)\frac{\partial H^{(\kappa)}}{\partial \pi_0}
\end{equation}
\end{subequations}
Here one particle Hamiltonian $H^{(\kappa)}$ is taken to be the
$\kappa$-invariant Casimir, Eq.\eqref{17a}:
$$H^{(\kappa)}\equiv\mu^2 =(\frac{\pi_0}{1-\pi_0})^2 -(\frac{\pi_i}{1-\pi_0})^2$$
If we use this expression in the Hamilton's equations (\ref{31a}),
(\ref{31b}) we obtain:
\begin{equation}\label{32}
  \frac{d\xi_i}{ds}= 2\frac{\pi_i}{(1-\pi_0)^2},
  ~~~~~~\frac{d\xi_0}{ds}= 2\frac{\pi_0}{(1-\pi_0)^2}
\end{equation}
From Eqs.(\ref{32}) immediately follows the expression for the
velocity (\ref{3a}) which was postulated from the very beginning:
\begin{equation}\label{33}
  \frac{d\xi_i}{d\xi_0}= \frac{\pi_i}{\pi_0}=v_i
\end{equation}\\

Another seemingly unphysical prediction(s) of Magueijo-Smolin (MS)
basis, as was pointed out by J.Rembielinski and K.Smolinski
\cite{JR1}, is connected with an apparent difficulty in
formulating statistical mechanics based on MS basis. It is argued
that one-particle partition function is divergent when $\pi_0
\rightarrow 1$. However this conclusion is based on an assumption
that the temperature in the Planck region is the same as in the
classical region. This is not true, since the existence of the
upper limit on the energy immediately implies that there exist a
relation between the temperature ( dimensionless) $\tau$ in the
Planck region and its counterpart $T$ in the classical region
analogous to the relations between energies in these two
regions,Eq.( \ref{11a}). As a result, it is not difficult to
demonstrate that the partition function (expressed as an integral) does not have singularities.\\

Additional criticism of $MS$ transformation is connected to the
fact that for a large number $N$ of identical particles, each of
energy ${\pi_0}_j, j=1,2,...,N$ their total internal energy  in
the thermodynamic limit ($N\rightarrow \infty$) does not depend on
temperature. But this represents not a deficiency of the basis,
but on the contrary, its advantage. In fact, since the temperature
is bounded from above by $\tau_{max}=1$, in the limit of infinite
number of particles, whose total energy tends to the respective
upper boundary ($\pi_0 =1$) the respective temperature $must$ tend
to its maximum that is to 1, which explains an apparent absence of
the dependence of the internal energy on temperature. In this case
the internal energy is simply equal to the temperature, and
both are equal to unity (in the chosen units).\\

In conclusion we would like to say that MS basis following from
very simple and consistent physical postulates introduced here
represents an attractive model for a description of phenomena
which might be associated with Planck scale physics. In fact, the
imposition of upper bound on the energy-momentum, and even mass (
if we adopt $\kappa$-Poinciana roots of the basis) which are in
agreement with a major postulate of Planck scale phenomena,  is
the feature which is $not~ present$ in any other bases. Still ,
there are some problems with this  ( and to this matter, with any
other $\kappa$-deformed) model(s). In particular, the commutation
relation $\{\pi_i,\xi_j\}=\delta_{ij}$ is not consistent with the
well-known string uncertainty relation.
\section{Addendum}
After this paper has been written, paper \cite{MS1} appeared,
where ( among some other topics) the addition law for
energy-momentum was modified as compared to the one used in
\cite{MV}. This was done to comply with the physical requirement
that a set of particles with even sub-Planckian energies can have
an energy much exceeding the Planck energy. The modification was
achieved by simply replacing the Planck energy $\kappa
(=1/\lambda$ in our units) for a system of
$N$ particles with $N\kappa$.\\

Here we demonstrate that the modified addition law follows from
our scheme, by simply adjusting one of our postulates (postulate
iii, \eqref{5}). The modification is as follows: we require that
for a set of $N$ particles the upper bound on both energy and
momentum ( normalized by Planck energy $\kappa$) is to be not $1$,
but $N$. This is equivalent to a postulate that the energy
composition law for particles ({\it each having Planck energy}) is
a simple addition. In turn, postulate ($iii$)for one particle
follows from that as a particular case of $N=1$.\\

Thus, if we use thus modified postulate in
Eqs.(\ref{11a},\ref{11b}), we obtain the following value of
constant $A$ which now depends on the number of particles $N$:
$$A(N)=\frac{1}{N},~~~ N=1,2,3,...$$
As a result, the energy $\pi_0^{(N)}$ and momentum $\pi_i^{(N)}$
of a set of $N$ particles follow from Eqs. (\ref{11a},\ref{11b})
(we restrict our attention to the positive values of $\pi_0$):
\begin{equation}\label{a1}
\pi_0^{(N)}= \frac{\Pi_0^{(N)}}{1+\Pi_0^{(N)}/N}
\end{equation}
\begin{equation}\label{a2}
\pi_i^{N}= \frac{\Pi_i^{(N)}}{1+\Pi_0^{(N)}/N}
\end{equation} The inverse expressions are
\begin{equation}\label{a3}
\Pi_0^{(N)}= \frac{\pi_0^{(N)}}{1-\pi_0^{(N)}/N}
\end{equation}
\begin{equation}\label{a4}
\Pi_i^{(N)}= \frac{\pi_i^{(N)}}{1-\pi_0^{(N)}/N}
\end{equation}
 $\Pi_0^{(N)}$ and $\Pi_i^{(N)}$ represent
the conventional sums of the respective individual quasi-energies
$\Pi_{0k}$ (momenta $\Pi_{ik}$) (as in special relativity):
\begin{equation}\label{a5}
\Pi_0^{(N)}=\sum_k^N \Pi_{0k},~~~~\Pi_i^{(N)}=\sum_k^N \Pi_{ik}
\end{equation}

Inserting \eqref{a5} into \eqref{a4},\eqref{a3}, we obtain the
composition law for energies $\pi_{0k}$ and momenta $\pi_{ik}$
\cite{MS1} which we write as follows:
\begin{eqnarray}\label{a6}
\pi_0^{N}= N\frac{\sum_{k=1}^N \pi_{0k}\prod_{j\neq
k}(1-\pi_{0j})}{\sum_{k=1}^N\prod_{j\neq
k}(1-\pi_{0j})(1-\delta_{jk})}
\end{eqnarray}\\
\begin{eqnarray}\label{a7}
\pi_i^{N}= N\frac{\sum_{k=1}^N \pi_{ik}\prod_{j\neq
k}(1-\pi_{0j})}{\sum_{k=1}^N\prod_{j\neq
k}(1-\pi_{0j})(1-\delta_{jk})}
\end{eqnarray}

The respective Casimir is found from Eqs.(\ref{a3}) and (\ref{a4})
if we take the mass $\mu$ to be defined the same way  in all the
regions, from classical to Planck's:
\begin{equation}\label{a8}
(\Pi_0^{(N)})^2-(\Pi^{(N)})^{2}=
(\frac{\pi_0^{(N)}}{1-\frac{\pi_0^{(N)}}{N}})^2-
(\frac{\pi_i^{(N)}}{1-\frac{\pi_0^{(N)}}{N}})^2=(\mu^{(N)})^2
\end{equation}
The composition laws Eqs.(\ref{a6}), (\ref{a7}) are reduced to the
conventional addition laws not only if all the individual energies
$\pi_{0k}$ (momenta $\pi_{ik}$)  are the same \cite{MS1}, but also
if at least one of the values $\pi_{0k}=1$ ($\pi_{ik}=1$). These
laws take especially simple form for two single particles:
\begin{eqnarray}
\pi_{01}\oplus\pi_{02}=2\frac{\pi_{01}+\pi_{02}-2\pi_{01}\pi_{02}}{2-\pi_{01}-\pi_{02}}\\
\pi_{k1}\oplus\pi_{k2}=2\frac{\pi_{k1}(1-\pi_{02})+\pi_{k2}(1-\pi_{01})}{2-\pi_{01}-\pi_{02}}
\end{eqnarray}

\end{document}